\documentclass[11pt,oneside,a4paper,reqno]{amsart}%
\pdfoutput=1
\usepackage[foot]{amsaddr}
\usepackage[version=3]{mhchem}
\usepackage{booktabs}
\usepackage[latin1]{inputenc}
\usepackage{upgreek}
\usepackage[pdftex]{graphicx}
\usepackage{amsmath}
\usepackage{amssymb}
\usepackage{microtype}
\usepackage{cite}
\usepackage{hyperref}

\usepackage{nicefrac}

\newcommand{\beginsupplement}{%
        \setcounter{table}{0}
        \renewcommand{\thetable}{S\arabic{table}}%
        \setcounter{figure}{0}
        \renewcommand{\thefigure}{S\arabic{figure}}
        \setcounter{equation}{0}
        \renewcommand{\theequation}{S\arabic{equation}}%
        \setcounter{page}{1}
        \renewcommand{\thepage}{S\arabic{page}}
        \setcounter{section}{1}
        \renewcommand{\thesection}{S\arabic{page}}
     }

\title[The Libration of Interfacial Water]{Experimentally Probing the Libration of Interfacial Water: the Rotational Potential of Water is Stiffer at the Air/Water Interface than in Bulk Liquid$^{*}$}

\author{Yujin Tong}
\address{Fritz Haber Institute of the Max Planck Society, Faradayweg 4-6, 14195 Berlin, Germany}
\author{Tobias Kampfrath}
\author{R.\ Kramer Campen}

\begin{document}

\maketitle
{\let\thefootnote\relax\footnotetext{$^{*}$ For the published version see, Tong et al (2016) Phys Chem Chem Phys, 18, 18424-18430 \href{http://dx.doi.org/10.1039/C6CP01004K}{(doi:10.1039/C6CP01004K)}}

\begin{abstract}
Most properties of liquid water are determined by its hydro\-gen-bond network. Because forming an aqueous interface requires termination of this network, one might expect the molecular level properties of interfacial water to markedly differ from water in bulk. Intriguingly, much prior experimental and theoretical work has found that, from the perspective of their time-averaged structure and picosecond structural dynamics, hydrogen-bonded OH groups at an air/water interface behave the same as hydrogen-bonded OH groups in bulk liquid water. Here we report the first experimental observation of interfacial water's libration (i.e.\ frustrated rotation) using the laser-based technique vibrational sum frequency spectroscopy. We find this mode has a frequency of 834 cm$^{-1}$, $\approx 165$ cm$^{-1}$ higher than in bulk liquid water at the same temperature and similar to bulk ice. Because libration frequency is proportional to the stiffness of water's rotational potential, this increase suggests that one effect of terminating bulk water's hydrogen bonding network at the air/water interface is retarding rotation of water around intact hydrogen bonds. Because in bulk liquid water the libration plays a key role in stabilizing reaction intermediates and dissipating excess vibrational energy, we expect the ability to probe this mode in interfacial water to open new perspectives on the kinetics of heterogeneous reactions at aqueous interfaces.
\end{abstract}
\section*{Introduction}

The chemical and physical properties of liquid water at the air/water interface control a wide variety of environmental, biological, and technological processes \cite{cac97,kni00,las03}. Many of the properties of \emph{bulk} liquid water, \textit{e.g.}\ its phase diagram, its density maximum in the liquid phase and its high specific heat capacity and viscosity relative to other liquids, are known to sensitively depend on water's hydrogen-bonding network. Since this network is terminated at the air/water interface, interfacial water molecules must be differently coordinated than those in bulk \cite{cla10}, one might expect that the properties of interfacial water differ significantly from those of the bulk liquid. Probing many macroscopic properties of the air/water interface, \textit{e.g.}\ surface tension and surface potential, is relatively straightforward. In contrast, it is challenging to experimentally probe the \emph{manner} in which water's hydrogen-bond network, and molecular-level structure and dynamics evolve as one moves from bulk to the air/water interface.   

Much work in hydrogen-bonded systems has shown that the OH stretch 
frequency sensitively reports on local environment \cite{ste02}. 
Unfortunately, applying conventional vibrational spectroscopies such as infrared absorption or spontaneous Raman scattering to 
probe interfacial water is generally impossible. Typically, the 
response of water molecules at the interface is overwhelmed by that of 
the much larger number in bulk. This challenge was first overcome by 
Shen and coworkers more than twenty years ago when they probed the OH stretch response of the 1-2 layers of water at the air/water interface 
\cite{du93} employing the interface-specific, laser-based technique, known as vibrational sum-frequency (VSF) spectroscopy. In a VSF measurement (see Figure \ref{f:scheme}a), a polarization is induced in water by a pulsed infrared (IR) field whose frequency is resonant with a molecular vibration. These oscillating dipoles then interact with an additional non-resonant field at visible frequencies (VIS) that upconverts only the emission from interfacial water molecules because they lack the local inversion symmetry of bulk\cite{she89}. 

Conducting a VSF measurement of the air/water interface with the IR incident field tuned to OH stretch frequencies gives an intensity spectrum with large features at $\approx$ 3150, 3400 and 3710 cm$^{-1}$ in IR frequency \cite{du93}. This spectrum differs dramatically from the OH stretch spectrum measured in IR absorption of bulk liquid water. Here, only a single peak is apparent that is centered at $\approx$ 3400 cm$^{-1}$ \cite{bak10}. The 3710 cm$^{-1}$ feature apparent in the VSF spectrum has been, uncontroversially, assigned to non-hydrogen-bonded OH groups that point towards the vapor phase. While the two lower-frequency features are known to correspond to OH groups that donate hydrogen bonds, understanding the manner in which they report on interfacial molecular structure and dynamics has proven challenging. Based on the similarity of the center frequencies of these features to those of the IR spectrum of ice and liquid water, a number of initial studies assigned the two peaks to quasi-static structural types of interfacial water: \textit{ice-like} and \textit{liquid-like} \cite{du93,du94,gra98,shu00}. However, more recent experimental and computational studies all appear to rule out the notion that these two peaks are the result of structural heterogeneity in interfacial water\cite{sov08a, sov08b,aue09,bon12,nih10, nih11, zha11,ino15}. While the appropriate assignment is still controversial, collecting VSF spectra of the OD stretch of HOD in H$_{\text{2}}$O at the air/water interface clarifies that the OH stretch line shape of hydrogen bonded interfacial water strongly resembles the OH stretch spectral response of the bulk liquid\cite{sov08a,sov08b,sov09,nih10,nih11}. Additionally, simulations of both time-averaged water structure, \textit{e.g.}\ hydrogen-bond distance (O-H$\cdots$O) and angle ($\angle$O-H$\cdots$O), and picosecond structural dynamics (\textit{i.e.}\ the change of OH groups to new hydrogen bond acceptors and the rotation of intact hydrogen bonded pairs \cite{laa08}) also find interfacial hydrogen-bonded OH groups behave similarly to those in bulk liquid water \cite{liu05, kuo06, kue11, aue09,vil12}.
\begin{figure}
	\includegraphics[width = 9 cm]{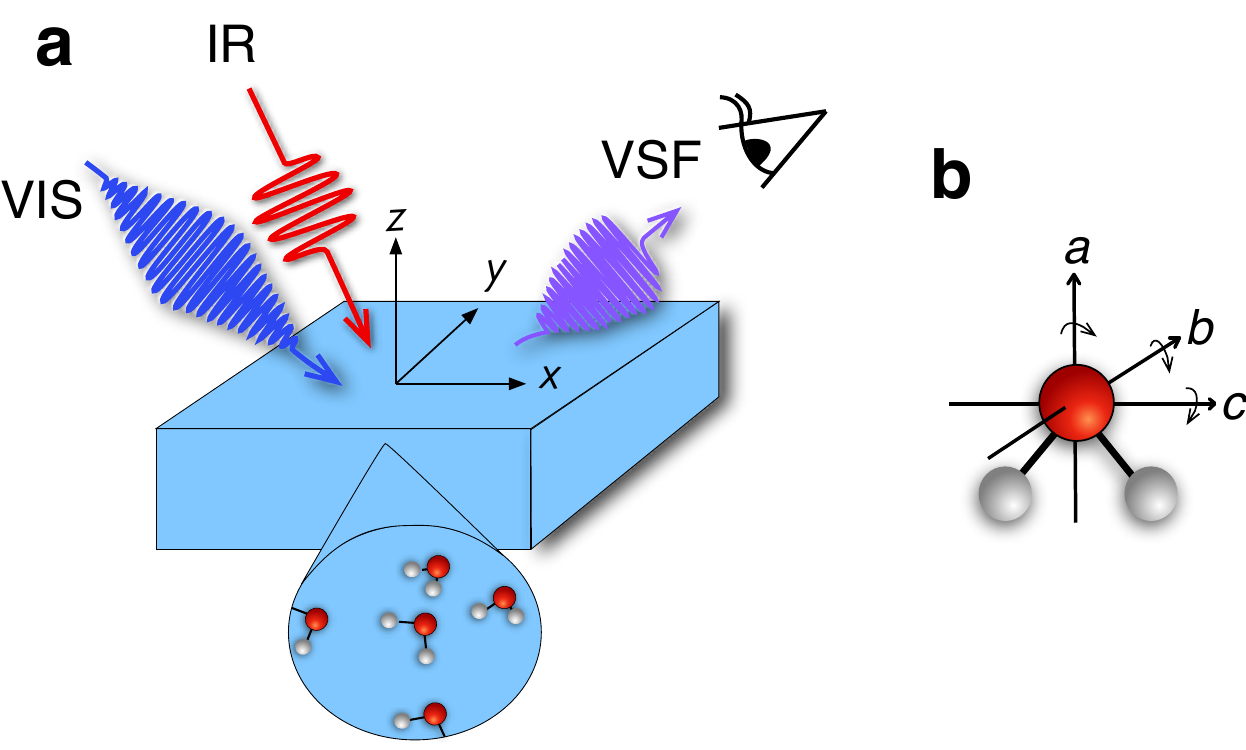}
	\caption{\label{f:scheme}(a) The experimental scheme for our VSF 
measurements. The infrared, visible and sum frequency beams all 
propagate in the \textit{x-z} plane. In geometry I the incident angles 
of the IR and VIS were set to 55$\pm$1$^{\circ}$ and 65$\pm$1$^{\circ}$ 
and in geometry II to 60$\pm$1$^{\circ}$ and 40$\pm$1$^{\circ}$. Light 
polarized in the \textit{x-z} plane is termed \textit{p} polarized, 
that perpendicular to this plane is termed \textit{s}. Following the 
usual VSF spectroscopy convention, a spectrum collected under the 
\textit{ssp} polarization condition indicates \textit{s}-polarized sum 
frequency, \textit{s}-polarized visible and \textit{p}-polarized 
infrared. (b) Coordinate scheme of an individual water molecule 
indicating the three possible librations.}
\end{figure}

These results are surprising. They suggest that terminating bulk water's hydrogen bond network has few consequences for hydrogen bond donating OH groups at an air/water interface. From an experimental point of view, this conclusion requires inferring the spatial arrangement of water molecules by their perturbation of an \emph{intramolecular} vibration: the OH stretch. In principle, probing \emph{intermolecular} vibrations is a more direct path to this insight. In bulk, the highest-frequency intermolecular vibration is water's libration (\textit{i.e.}\ frustrated rotation): the frequency of which is proportional to the stiffness of the potential of mean force describing water's rotation about intact hydrogen bonds (see Figure \ref{f:scheme}b). As noted above, from a simulation point of view most studies of the air/water interface have focussed on either time-averaged structure or picosecond structural dynamics. Such long timescale dynamics reflect changes in network topology, hydrogen bonds break and reform on ps timescales, but do not tell us whether the stiffness of the intact hydrogen bonded network changes as one moves from bulk to the interface.

Probing the libration of \emph{interfacial} water addresses much of the limitations of prior experiment and simulation. From an experimental perspective it clearly complements and extends prior work characterising the OH stretch of interfacial water by more directly addressing water's intermolecular potential. From the perspective of simulation it complements the prior work described above by directly addressing the properties of the intact hydrogen-bonded network at the air/water interface. In addition to the insight it offers into water structure and dynamics a deeper understanding of water's libration at interfaces is of interest in its own right. Much prior work in bulk has shown that this mode plays an essential role in the flow of excess vibrational energy (\textit{e.g.}\ from excited OH stretch modes to heat \cite{ash06,ing09,pet13}) and in many aqueous phase chemical reactions (\textit{e.g.}\ the localization of the excess electron in water \cite{gar05,zhu13b,sav14}). The ability to probe the libration at aqueous \emph{interfaces} should allow significant insight into the kinetics and mechanisms of these and similar processes in heterogeneous aqueous phase chemistry \cite{kni00,las03,jun07,abe13}.

Here, then, we employ VSF spectroscopy to experimentally characterize, for the first time, the libration of interfacial water. We find that the libration frequency of interfacial water is significantly higher than bulk liquid water at the same temperature. This observation suggests water's rotational potential stiffens on moving from the bulk liquid to the air/water interface.

\section*{Results and Discussion}
\subsection*{Experimental Observable}
To do so, we constructed the experiment shown schematically in Figure \ref{f:scheme} and collected a VSF spectrum in the IR frequency range 680-1050 cm$^{-1}$. The technical details of our VSF spectrometer have  been described previously\cite{ton15}, and are reproduced here, along with details specific to the current study, in the Electronic Supplementary Information. The results of this measurement are shown in Figure \ref{f:sfg_data}a. Note that we have plotted the intensity of the emitted VSF light ($I_{\text{VSF}}$) divided by the intensity of the incident infrared ($I_{\text{IR}}$) to account for the fact that the power of our incident infrared field is frequency dependent. Clearly, this spectrum shows a striking feature centered at $\approx 875$ cm$^{-1}$. To confirm this feature is a property of interfacial water, we performed a variety of control experiments. To verify the surface was contaminant free, we collected VSF spectra in this frequency range at the air/\ce{D2O}, and CH and OH stretch frequency ranges at the air/\ce{H2O} interface. To verify the signal was, in fact, sum-frequency emission, we confirmed that it was emitted at the angle expected from wave-vector conservation \cite{lam05}, and that $I_{\text{VSF}}$ depended linearly on the intensity of the VIS and IR beams. All control experiments thus strongly suggest that the spectral feature we observe is a property of interfacial \ce{H2O} (see section 1 of the Electronic Supplementary Information for details).

\begin{figure}
\begin{center}
	\includegraphics[width = 10 cm]{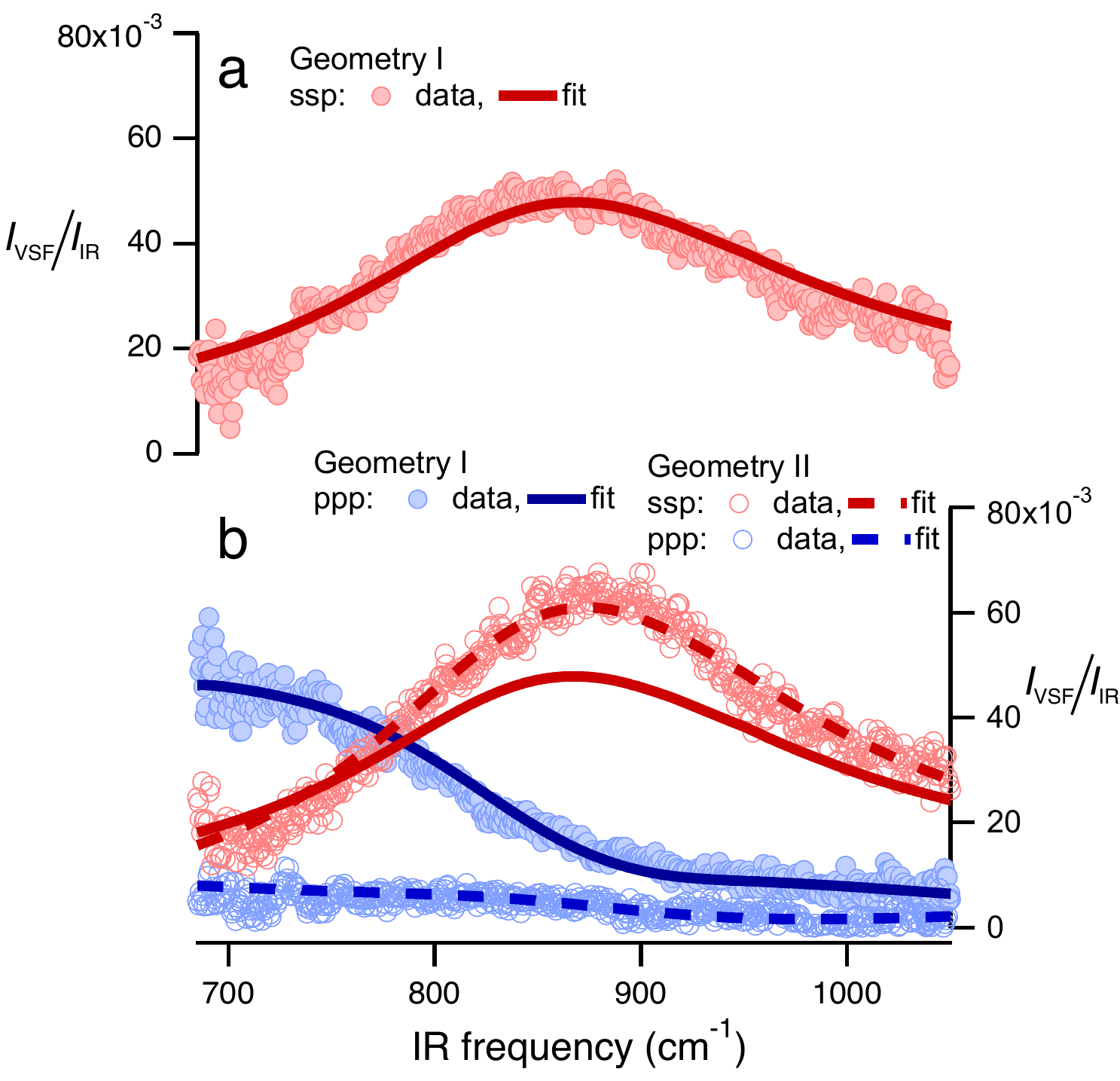}
	\caption{$\nicefrac{I_{\text{VSF}}}{I_{\text{IR}}}$ plotted as a function of incident IR frequency for spectra collected in (\textbf{a}) geometry I under the \textit{ssp} polarization condition and (\textbf{b}) in geometry I under the \textit{ppp}, and geometry II under both \textit{ssp} and \textit{ppp}. Solid and dashed lines are fits to the data using the line-shape model described in the text. The fit of the spectra shown in (a) is reproduced in (b) for comparison. All spectra shown have been normalized to account for the frequency-dependent IR intensity and to account for intensity variations between experimental configurations (see section 2 of the Supplementary Information for details).}
	\label{f:sfg_data}
\end{center}
\end{figure}
Note that the $\nicefrac{I_{\text{VSF}}}{I_{\text{IR}}}$ spectrum shown in Figure \ref{f:sfg_data}a contains not only the vibrational response of interfacial water molecules but also depends on the incident angles and polarizations of all beams. The interfacial vibrational information is  contained in the dependence of interfacial water's nonlinear susceptibility ($\chi^{(2)}$) on the frequency of the incident infrared field. As is shown in equation (\ref{e:sig_tot}), however, in addition to $\chi^{(2)}$, $\nicefrac{I_{\text{VSF}}}{I_{\text{IR}}}$ depends on the well-known Fresnel coefficients (\textbf{\textit{L}}) describing the reflection and transmission of all fields at an interface and their polarizations ($\hat{e}$ is the unit polarization vector), the spectrum of the visible pulse ($E_{\text{VIS}}$, $\otimes$ indicates a convolution), and the emission angle of the VSF field ($\theta_{\text{VSF}}$)  \cite{zhu99,wan05,ton13}.
\begin{multline}\label{e:sig_tot}
%
		\frac{I_{\text{VSF}}(\nu_{\text{IR}} + \nu_{\text{VIS}})}
{I_{\text{IR}}(\nu_{\text{IR}})} \propto \sec^{2}\theta_{\text{VSF}} \\
\left|E_{\text{VIS}}\otimes \left([\hat{e}_{\text{VSF}}\cdot\textbf{\textit{L}}_{\text{VSF}}]\cdot\chi^{(2)} 
(\nu_{\text{IR}}): [\hat{e}_{\text{VIS}}\cdot\textbf{\textit{L}}_{\text{VIS}}][\hat{e}_{\text{IR}}\cdot\textbf{\textit{L}}_{\text{IR}}]\right)\right|^{2}
%
\end{multline}

The Fresnel coefficients depend on the incident angles of the fields, their polarizations and the frequency-dependent, refractive indices of water and air. Their functional form has been well described previously and is reproduced, along with the complete expressions for the reflected VSF signal, in section 2 of the Electronic Supplementary Information \cite{zhu99, wan05, ton13}. For purposes of this study the key point is that they are known \emph{independently} of our measurements. Thus, calculating the Fresnel coefficients and independently measuring the spectrum of our visible pulse, \textit{i.e.}\ $E_{\text{VIS}}$, we can extract $\chi^{(2)}$ from the data. 

We quantify $\chi^{(2)}$ by following prior workers and assume it is a coherent superposition of a nonresonant contribution ($\chi^{(2)}_{\text{nr}}$), and one or more 
Lorentz-type resonances ($\chi^{(2)}_{\text{r}}$) \cite{bai91}: 
\begin{eqnarray}\label{e:line}
\chi^{(2)} & = & \chi^{(2)}_{\text{nr}} + \chi^{(2)}_{\text{r}} 
\nonumber \\
	& = & \left| \chi_{\text{nr}} \right|\text{e}^{\text{i}\epsilon} + 
\sum_{n}\frac{\chi_{n}}{\nu_{\text{IR}} - \nu_{n} + \text{i}
\Gamma_{n} } 
\end{eqnarray}
in which $\left| \chi_{\text{nr}} \right|$ and $\epsilon$ are the nonresonant 
amplitude and phase, and $\chi_{n}$, $\nu_{n}$ and $\Gamma_{n}$ are 
the complex amplitude, center frequency and damping constant of the 
$n^{\text{th}}$ resonance. We model our data by combining equation 
(\ref{e:line}) with the appropriate Fresnel coefficients and substituting the resulting product into equation (\ref{e:sig_tot}) (see section 2 of the Electronic Supplementary Information for full details). The resulting fit to the data using this approach and a single resonance is shown in Figure \ref{f:sfg_data}. 

Because water's reflectivity is strongly frequency-dependent over 680-1050 cm$^{-1}$, one would expect a dramatic change in measured $\nicefrac{I_{\text{VSF}}}{I_{IR}}$ as a function of angles and polarizations of the incident beams. If our line-shape model is correct, these changes should be entirely captured in the Fresnel coefficients: data collected under different incident beam angles and polarizations should be well described by the same $\chi^{(2)}$ (\textit{i.e.}\ the same parameters in equation (\ref{e:line})). As shown in Figures \ref{f:sfg_data}a and \ref{f:sfg_data}b, we find that measuring our sample with different beam incident angles and polarizations leads to drastic changes in $\nicefrac{I_{\text{VSF}}}{I_{\text{IR}}}$ but that all spectra can be well described with a single resonance with the same center frequency, line width and amplitude: the drastic differences in the measured spectra are quantitatively described by changes in Fresnel coefficients. The center frequency of the resulting resonance is 834 cm$^{-1}$. We emphasize that assuming $\chi^{(2)}_{r}=0$ results in a qualitative misfit of the data (see section 4.1 of the Electronic Supplementary Information for more details).

\subsection*{Assignment}
Having demonstrated that we probe an 834 cm$^{\text{-1}}$ resonance of interfacial water, we turn to the vibrational response of the bulk liquid to help in this mode's assignment. Water's libration is the closest in frequency to our observed feature (in bulk liquid water no other mode is within a factor of two in energy \cite{mar11b}). Probed via IR absorption, in bulk liquid water it has a center frequency of $\approx$ 670 cm$^{\text{-1}}$ at 25 $^{\circ}$C. On cooling, it narrows and blue-shifts by $\approx$ 165 cm$^{\text{-1}}$ until the formation of ice \cite{zel95,ber96,sev03,pet13}. Probed via spontaneous Raman scattering, the spectral response of water's libration is more complicated. At 25 $^{\circ}$C, modes are apparent at 450, 550, and 722 cm$^{-1}$ \cite{wal67} (where the 450 cm $^{-1}$ is IR inactive). However, the quantitative trend with cooling is the same: spectral weight shifts to higher frequencies with cooling until reaching ice where modes are apparent at 682, 799 and 920 cm $^{\text{-1}}$ \cite{ben87,won75}. As shown in Figure \ref{f:sfg_temp} we find that our observed spectral weight blue-shifts on cooling from 23 to 0 $^{\circ}$C, similar to both bulk-sensitive techniques. 
\begin{figure}
\begin{center}
	\includegraphics[width = 9 cm]{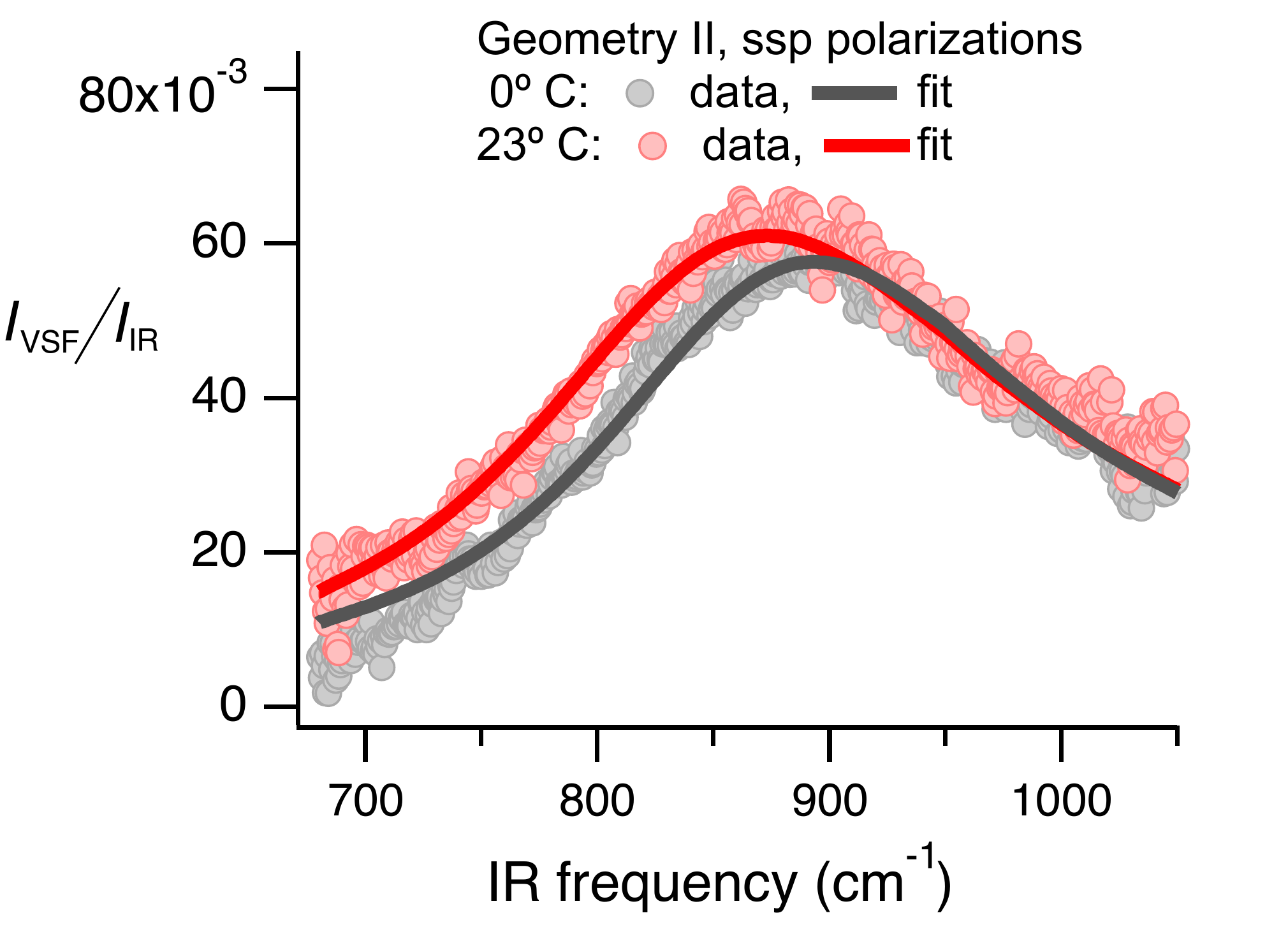}
	\caption{$\nicefrac{I_{\text{VSF}}}{I_{\text{IR}}}$ as a function of infrared frequency collected under the \textit{ssp} polarization condition at 0$^{\circ}$ and 23$^{\circ}$C.}
	\label{f:sfg_temp}
\end{center}
\end{figure}

While there are no prior VSF measurements of the libration at liquid 
water interfaces, two published computational studies have reported VSF 
spectra of the air/water interface at these frequencies. Both Perry et al.\ and Nagata et al.\ find a single broad peak, qualitatively similar to that of the IR absorption in bulk water \cite{per05b,nag13}. Quantitatively, however, the results of the two studies differ. Perry et al.'s findings agree remarkably well with our measured response: the librational mode of water at the air/water interface is blue-shifted $\approx$ 200 cm$^{-1}$ relative to that of bulk water \cite{per05b}. Similar to our results (see Figure \ref{f:sfg_data}) they predict the resonant intensity observed under the \textit{ssp} polarization condition to be $\approx 4\times$ larger than that under \textit{ppp}. In contrast, Nagata and coworkers, as part of a study on the bend of water at the air/water interface, found the interfacial librational spectral response quantitatively similar to bulk \cite{nag13}. However, their sensitivity tests and prior studies by others \cite{hey12} strongly suggest that the spatial truncation scheme adapted to maximize computational efficiency, while suitable for study of the bending mode of interfacial water, is not appropriate for accurate description of the libration.

To summarise, we have demonstrated that the feature apparent 
at 875 cm$^{-1}$ in our VSF spectra is the result of a resonance of interfacial water centered at 834 cm$^{-1}$. Based on (i) the similarity in the center frequency of $\chi^{(2)}_{r}$ to that of the libration of bulk water measured via IR absorption and spontaneous Raman scattering, (ii) the similar temperature dependence of the resonance we observe and bulk water's libration, and (iii) the close correspondence between our measured signal and that predicted for the libration of interfacial water by Perry et al.\ \cite{per05b}, we assign the 834 cm$^{-1}$ resonance to the libration of water at the air/water interface.

Before moving onto the implications of this observation it is worth noting that in bulk water the libration of \ce{D2O} is shifted by $\approx$ 170 cm$^{-1}$ to lower energies than that of \ce{H2O} \cite{zel95}. Should the feature we observe be the libration of interfacial \ce{H2O}, and the change of libration frequency when changing from \ce{H2O} to \ce{D2O} be similar in bulk liquid water and at the air/water interface, we would expect that probing the air/\ce{D2O} interface in the 700-1050 cm$^{-1}$ should reveal only the high frequency shoulder of a mode centered at lower frequencies. Our measurements (see Section 3 of the Electronic Supplementary Information) are consistent with this expectation, further suggesting that the resonance we observe at the air/\ce{H2O} interface is the libration of interfacial water.

\subsection*{Implications for Interfacial Water Structure}
The fact that the libration changes frequency as one moves from bulk liquid to the air/water interface while the OH stretch, and time-averaged structure and ps structural dynamics extracted from simulation, do not, suggests that this resonance provides new insight into interfacial water structure. Because the existence of a libration in liquid water is a consequence of the hindering of water's rotation by hydrogen bonds, the libration frequency is a measure of the ease of changing hydrogen bond angles for an intact hydrogen bond. Alternatively, the libration frequency is proportional to the stiffness of water's rotational potential while hydrogen bonded. To understand why water's libration at the air/water interface should differ significantly in frequency from that of the same mode in the bulk liquid, but many other observables differ little between these two environments, we first review work developing a microscopic picture of how the OH stretch relates to water structure.

As emphasized by Geissler in a recent review \cite{gei13}, and following work by a number of other groups \cite{sad99,cor04, eav05, hay05,pae09}, the OH stretch frequency of water molecules decreases on moving from gas to liquid due to coupling to low-frequency modes and the magnitude of this change is proportional to the local electric field on an OH group's hydrogen atom projected in the direction of the OH bond. The change in OH stretch frequency (\textit{i.e.}\ $\Delta\nu_{\text{OH}}^{\text{g}\rightarrow\text{l}}$) is thus quite sensitive to the hydrogen-bond distance ($\Delta\nu_{\text{OH}}^{\text{g}\rightarrow\text{l}} \propto$1/(O-H$\cdots$O)$^{2}$), much less sensitive to the hydrogen-bond angle ($\Delta\nu_{\text{OH}}^{\text{g}\rightarrow\text{l}} \propto (\cos\angle$O-H$\cdots$O)), and relatively insensitive to the presence of additional water molecules, \textit{i.e.}\ three-body and higher interactions.  In this simplified picture, the linewidth of the OH stretch spectral response is the result of static disorder: OH groups do not change frequency on the time scale of light absorption. Calculations of the full VSF spectral response suggest this approximation is surprisingly accurate, accounting for motional narrowing is not necessary to qualitatively reproduce experimental VSF spectra \cite{aue08}. Thus, the OH stretch frequency reports principally on hydrogen-bond distance and a decrease in frequency indicates a decrease in distance. 

Because VSF studies of the OH stretch at the air/water interface (when studying isotopically diluted H$_{\text{2}}$O) find a center frequency of the OH stretch similar to bulk liquid water \cite{sov09}, this logic implies that hydrogen bond distances should  be the same in both environments. Indeed simulations of average hydrogen bond distance and angle at the air/water interface confirm this picture \cite{vil12}. In addition to studying time averaged structural properties, atomistic simulation can be straightforwardly used to study the structural dynamics of water. Such studies in bulk have clarified that water's hydrogen bond network breaks and reforms rapidly, \textit{i.e.}\ transfer of a hydrogen bond from an old to a new acceptor, so-called hydrogen bond exchange, takes $\approx$ 100 fs, but infrequently, \textit{i.e.}\ once every several ps, and involves the concerted motion of $>3$ water molecules \cite{liu05,laa08}. Similar work at the air/water interface has shown that the mechanism by which hydrogen bonds break and reform is quite similar to bulk liquid and that such reforming of water's hydrogen bond network also occurs on ps timescales \cite{vil12}. While it thus seems clear that changes in libration frequency need not be related to changes in slower structural dynamics, it is less clear that there should be no relation between hydrogen bond distance and the stiffness of water's rotational potential. Certainly at least both of these observables describe the hydrogen-bonded state.

Computational studies of the gas-phase water dimer make clear, \textit{e.g}\ see Scott and Vanderkoi \cite{sco10}, that in the gas phase as the two waters approach each other the OH rotational potential is stiffened. As a result OH stretch frequency and libration frequency are strongly anti-correlated: with decreasing hydrogen bond distance OH stretch frequency decreases and libration frequency increases. For a gas phase water dimer the most energetically favorable hydrogen bonding distance and angle depends only on the relative position of the two molecules. It is not immediately obvious that such a 2-body character describes the intermolecular potential in, and thus that such a OH stretch/libration frequency correlation should extend to, liquid water.

X-ray, neutron and thermodynamic studies of bulk liquid water demonstrate that with decreasing temperature hydrogen bonds increase in strength and hydrogen bond distance decreases \cite{laz96,hur00,sop00}. Consistent with these results, with decreasing temperature OH stretch frequency shifts to lower energies \cite{mar11b}. As discussed above, and also consistent with the trend expected from the dimer, with decreasing temperature the libration frequency shifts to higher energies \cite{zel95}. That is, with stronger hydrogen bonding in liquid water hydrogen-bond distance shortens and the rotational potential stiffens. While the correlation of the temperature-dependent changes in libration and OH stretch frequency of liquid water thus resemble those of the dimer, intriguingly, Fayer and coworkers have demonstrated that at low temperatures the 2-body character of the intermolecular potential does not \cite{moi08}. In particular, they show that with cooling, \textit{i.e.}\ stronger hydrogen bonding, water's rotational potential increasingly changes from a 2-body to multi-body character \cite{moi08}. This change is rationalized by noting that water's structural correlations also increase dramatically with decreasing temperature \cite{moo09,nil12}. Consistent with the logic of Geissler and coworkers described above, OH stretch frequencies are relatively insensitive to this change in character of water's intermolecular potential. Fayer and coworkers observe it only via a temperature dependent \textit{decoupling} of the librational contribution to the OH orientational correlation function and OH stretch frequency.     

Given this understanding of bulk liquid water it seems reasonable to ask whether an air/water interface also induces structural correlations. Simulation studies clarify that, while hydrogen bond distances are quite similar at the air/water interface and in bulk liquid water, structural correlations are dramatically enhanced (this is immediately evident from the fact that interfacial water molecules have a preferential orientation with respect to an external reference frame while those in bulk do not) \cite{tay96,vil12}. Armed with both of these observations it is perhaps unsurprising that when moving from bulk liquid water to the air/water interface the OH stretch frequency is constant but the libration frequency is not. Evidently, forming the air/interface induces structural correlations similar to those found when cooling liquid water without the decrease in hydrogen bond distance. 

It is finally worth noting that, while the frequency and polarization  dependence of our measured interfacial libration are well predicted by one of the two published studies of the librational response of VSF spectra of water at the air/water interface, much work clearly remains to be done. For example, while both theory studies find that the libration of water at the air/water interface, similar to bulk IR absorption measurements, is only a single spectral feature \cite{per05b,nag13}, we cannot rule out the possibility that there is a second librational resonance of interfacial water at frequencies $< \text{650 cm}^{-1}$ (\textit{i.e.}\ outside the detection window of our current experimental set up). Similarly, while our data are well described by a single, homogeneously broadened resonance, we cannot rule out the possibility of inhomogeneous broadening due either to librations of double hydrogen-bond donors at multiple and distinct frequencies, distinct librations of singly hydrogen-bond donating species or other forms of structural heterogeneity \cite{wer04,wan05, hua09,tas13,hsi14,tas14}. We also cannot rule out the possibility such populations exist and do not contribute to our observed VSF signal due to orientational effects. From an experimental point of view, it is clear that extending our VSF measurements to significantly lower IR frequencies, \textit{e.g.}\ 400 cm$^{-1}$, and conducting 2D IR/VSF measurements over this frequency range would substantially constrain much of this uncertainty. Additionally quantifying the OH stretch rotational anisotopy at the air/water interface in a frequency and polarisation resolved IR pump / VSF probe scheme, extending the experimental concept of Fayer and coworkers to interfaces, would directly address whether the character of the intermolecular potential at the air/water interface differs from that in bulk at the same temperature \cite{moi08,hsi11}. All of these measurements are technically demanding and the object of current work in our group. Despite these limitations, our current study offers a new observable describing the structure and dynamics of water at the air/water interface.

\section*{Conclusions}
The observation that the frequency of the libration of interfacial water is blue-shifted by 165 cm$^{-1}$ relative to that in bulk liquid water suggests that the potential of mean force underlying water rotation is significantly stiffer at the air/water interface than in bulk. In contrast, much prior experiment and simulation investigating the time averaged structure and picosecond structural dynamics of hydrogen-bonded OH groups at the air/water interface find that, from the perspective of these observables, \textit{interfacial water is similar to bulk liquid water}. There is no contradiction between these apparently discordant viewpoints. Water's OH stretch frequency is principally sensitive to hydrogen bond distance while libration frequency is proportional to the stiffness of water's rotational potential around intact hydrogen bonds: each observable samples different aspects of water's intermolecular potential. Similarly, water's picosecond structural dynamics involve the breaking and reforming of hydrogen bonds and require concerted motions of $>3$ water molecules. These motions are substantially less frequent than those of water's rotational libration, they occur at THz frequencies, and are thus not sampled in our spectral window.   

In this study we describe a novel probe of the intact hydrogen bond network of interfacial water: interfacial water's libration. In addition to the insight into water physics it offers, the ability to probe interfacial water's libration immediately suggests a number of intriguing avenues of future research. As has been well documented in bulk, the water's libration is both important in understanding the dissipation of excess vibrational energy \cite{ash06,ing09,pet13} and in many aqueous phase chemical reactions. Probing the \emph{interfacial} libration makes it possible to gain similar mechanistic insights into vibrational relaxation and heterogeneous chemistry at aqueous interfaces \cite{kni00,las03,jun07,abe13}.

\section*{Electronic Supplementary Information}
\noindent Full description of the theoretical expressions of the reflected VSF signal, the procedure to normalize the signal to account for the spectral profile of the incident IR field, the procedure to quantitatively compare between the different experimental geometries, a description of the control experiments and a description of the details of the quantitative line shape analysis.

\clearpage

\beginsupplement

\section{Electronic Supplementary Information}

\subsection{Details of our VSF Spectrometer and Sample}
The VSF spectrometer used in this work consists of a mode-locked Ti:Sapphire Oscillator (Venteon), a regenerative amplifier and a subsequent cryo-cooled amplifier (Coherent). The amplifier delivers 800 nm, 35 fs, 15 mJ pulses at a repetition rate of 1 kHz. Half of this energy is used to generate infrared pulses centered at 12 $\mu$m ($\approx$ 833 cm$^{-1}$) by difference frequency mixing of the signal and idler output of a commercial optical parametric amplifier (TOPAS, Light Conversion). The residual of the 800 nm light from the parametric amplification is spectrally narrowed using a home-made pulse shaper -- composed of a grating, mirror and slit combination -- and used as the up-conversion field in the sum frequency process (\textit{i.e.}\ the \textit{visible} field in the usual VSF terminology). The spectral width of the visible pulse is set to $<$ 2 nm. For all measurements the pulse energy of the infrared at the sample surface was 6.0 and that of the visible 18 $\mu$J.  The infrared and visible beams were focused on the samples by lenses with focal lengths of 10 and 35 cm, and, as noted in the manuscript, had incident angles of 55$\pm$1$^{\circ}$ and 65$\pm$1$^{\circ}$ for geometry I and 60$\pm$1$^{\circ}$ and 40$\pm$1$^{\circ}$ for geometry II. The size of the infrared and visible foci at the sample were 80 and 200 $\mu$m respectively.  After the sample, the VSF signal was collimated by a lens (f=450 mm) and focused again on the entrance slit of a spectrograph (Andor technology, Shamrock 303). For detection the dispersed signals were imaged on an EMCCD camera (Andor, Newton).

We calibrated our spectrograph using emission lines from a Neon lamp; the frequency calibration of the spectrum was performed using a polystyrene thin film and relative VSF intensities in different experimental geometries were quantitatively compared using the resonant signal from bulk z-cut quartz as a reference. To account for the frequency-dependent IR pulse energy, we normalized the measured VSF response from the air/\ce{H2O} interface by the, nonresonant, VSF signal from a gold mirror.  The acquisition time for measurements of the VSF spectral response of the gold mirror and \ce{H2O} interface were 30 and 600 seconds, respectively.

H$_{2}$O measurements were performed in a home built teflon trough using Millipore MilliQ (resistivity $> 18.2 \text{ M}\Omega\cdot\text{cm}$) water.

\subsection{Quantitatively Comparing I$_{\text{VSF}}$ Collected in Different Experimental Geometries}
\subsubsection{Quantifying Reflected I$_{\text{VSF}}$}
As discussed in detail in prior work \cite{zhu99SI,wan05SI,ton13SI}, given an interface irradiated by optical fields at infrared ($E_{\text{IR}}$) and visible ($E_{\text{VIS}}$) frequencies, and describing the system in the electric dipole approximation, a second order nonlinear polarization is generated in the interfacial layer that radiates in the reflected direction, 
	\begin{equation}\label{e:sig_tot_SI}
		I_{\text{VSF}} = \frac{ 8\pi\nu^{3}_{\text{VSF}}\sec^{2}\theta_{\text{VSF}}}{c^{3}}\left|E_{\text{VIS}}(\nu_{\text{VIS}})\otimes \chi^{(2)}_{\text{eff}} \right|^{2}I_{\text{IR}}(\nu_{\text{IR}})
	\end{equation}
in which $E_{\text{VIS}}$ is the, frequency dependent, 800 nm field, $\otimes$ is the convolution operator, $\chi^{(2)}_{\text{eff}}$ is the effective macroscopic second order nonlinear susceptibility, $\theta_{\text{VSF}}$ is the angle at which the sum frequency emission radiates and $I_{\text{IR}}$ is the, frequency dependent, infrared intensity.

$\chi^{(2)}_{\text{eff}}$ is a function of the second-order optical response of the material $\chi^{(2)}$, the polarizations of the incident and outgoing fields, and the Fresnel coefficients. It can be expressed as (in which $\hat{e}$ is the unit polarization vector of the indicated beams), 
\begin{equation}\label{e:chi2_eff}
	\chi^{(2}_{\text{eff}} = \left[\hat{e}_{\text{VSF}}\cdot\textbf{\textit{L}}_{\text{VSF}}\right]\cdot\chi^{(2)}(\nu_{IR}):[\hat{e}_{\text{VIS}}\cdot\textbf{\textit{L}}_{\text{VIS}}][\hat{e}_{\text{IR}}\cdot\textbf{\textit{L}}_{\text{IR}}]
\end{equation}  
As described in the manuscript, we fit the data by substituting equation (\ref{e:chi2_eff}) into (\ref{e:sig_tot_SI}).

For interfaces, such as the air/water, that have macroscopic $C_{\infty\nu}$ symmetry, there are seven nonzero and four independent components of $\chi^{(2)}$. Given the experimental geometry shown in the manuscript, z is along the surface normal and all fields propagate in the x-z plane, these nonzero components are, $\chi^{(2)}_{xxz} = \chi^{(2)}_{yyz}$, $\chi^{(2)}_{xzx} = \chi^{(2)}_{yzy}$, $\chi^{(2)}_{zxx} = \chi^{(2)}_{zyy}$ and $\chi^{(2)}_{zzz}$. Measuring $I_{\text{VSF}}$ under the \textit{ssp} and \textit{ppp} polarization conditions (in which \textit{s} indicates a field perpendicular to the x-z plane and \textit{p} parallel) samples these nonzero components of the $\chi^{(2)}$,
\small{
\begin{eqnarray}
\chi^{(2)}_{\text{eff},ssp} & = & L_{yy}(\nu_{\text{\tiny{VSF}}})L_{yy}(\nu_{\text{\tiny{VIS}}})L_{zz}(\nu_{\text{\tiny{IR}}})\sin\theta_{\text{\tiny{IR}}}\chi^{(2)}_{yyz}\\
\chi^{(2)}_{\text{eff},ppp} & = & -L_{xx}(\nu_{\text{\tiny{VSF}}})L_{xx}(\nu_{\text{\tiny{VIS}}})L_{zz}(\nu_{\text{\tiny{IR}}})\cos\theta_{\text{\tiny{VSF}}}\cos\theta_{\text{\tiny{VIS}}}\sin\theta_{\text{\tiny{IR}}}\chi^{(2)}_{xxz}\nonumber\\
& & -L_{xx}(\nu_{\text{\tiny{VSF}}})L_{zz}(\nu_{\text{\tiny{VIS}}})L_{xx}(\nu_{\text{\tiny{IR}}})\cos\theta_{\text{\tiny{VSF}}}\sin\theta_{\text{\tiny{VIS}}}\cos\theta_{\text{\tiny{IR}}}\chi^{(2)}_{xzx}\nonumber\\
& & +L_{zz}(\nu_{\text{\tiny{VSF}}})L_{xx}(\nu_{\text{\tiny{VIS}}})L_{xx}(\nu_{\text{\tiny{IR}}})\sin\theta_{\text{\tiny{VSF}}}\cos\theta_{\text{\tiny{VIS}}}\cos\theta_{\text{\tiny{IR}}}\chi^{(2)}_{zxx}\\
& &+L_{zz}(\nu_{\text{\tiny{VSF}}})L_{zz}(\nu_{\text{\tiny{VIS}}})L_{zz}(\nu_{\text{\tiny{IR}}})\sin\theta_{\text{\tiny{VSF}}}\sin\theta_{\text{\tiny{VIS}}}\sin\theta_{\text{\tiny{IR}}}\chi^{(2)}_{zzz}\nonumber
\end{eqnarray}
}
in which $\theta_{i}$ is the angle of the i$^{th}$ field with respect to the surface normal and $L_{jj}(\Omega)$ are the diagonal elements of the Fresnel factors evaluated at frequency $\Omega$. These diagonal elements are,
\begin{eqnarray}
L_{xx}(\Omega) & = & \frac{2\cos\gamma}{\cos\gamma + n_{\text{H$_{2}$O}}(\Omega)\cos\theta_{i}}\label{e:f_x} \\
L_{yy}(\Omega) & = & \frac{2\cos\theta_{i}}{\cos\theta_{i} + n_{\text{H$_{2}$O}}(\Omega)\cos\gamma} \label{e:f_y} \\
L_{zz}(\Omega) & = & \frac{2n_{\text{H$_{2}$O}}(\Omega)\cos\theta_{i}}{\cos\gamma + n_{\text{H$_{2}$O}}(\Omega)\cos\theta_{i}} \left(\frac{1}{n^{\prime}(\Omega)}\right)^{2}\label{e:f_z}
\end{eqnarray}
in which $n_{\text{H$_{2}$O}}(\Omega)$ is the refractive index of water at frequency $\Omega$, $\theta_i$ is the angle of incidence of the $i^{th}$ beam (either IR, VIS or VSF), $\gamma$ is the refraction angle (\textit{i.e.} $\sin\theta_{i} = n_{\text{H$_{2}$O}}(\Omega)\sin\gamma$) and $n^{\prime}(\Omega)$ is the interfacial refractive index at frequency $\Omega$. The interfacial refractive index for this system is calculated using the approach described by Shen and coworkers \cite{zhu99SI}. 
\begin{figure}
\begin{center}
	$\begin{array}{c c}
	\includegraphics[width=0.5\textwidth]{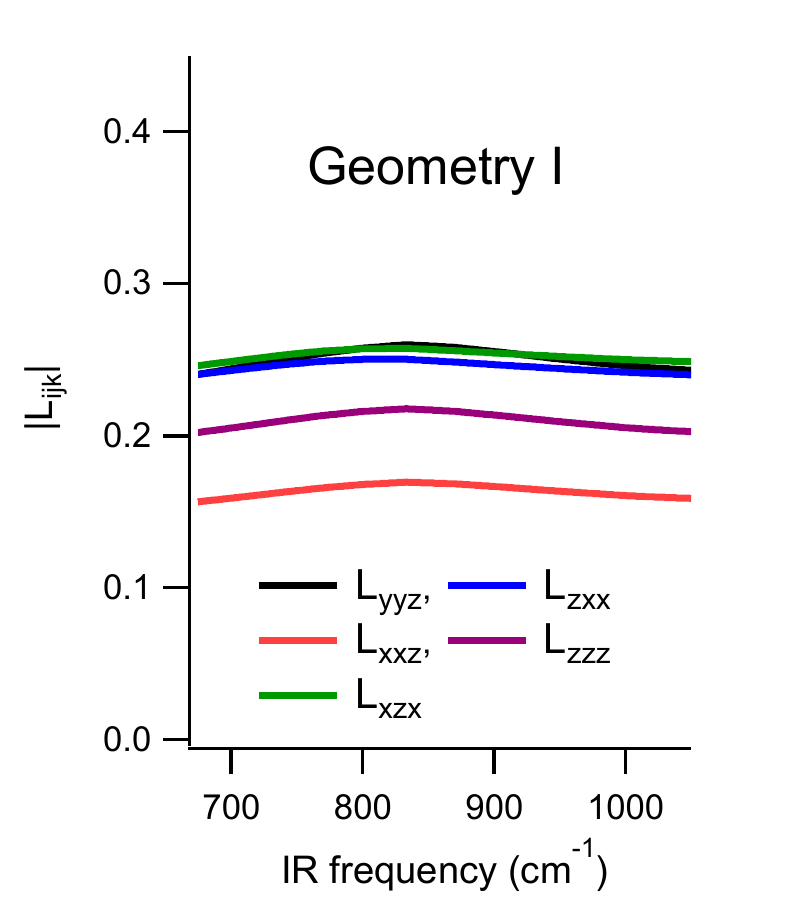} &
	\includegraphics[width=0.5\textwidth]{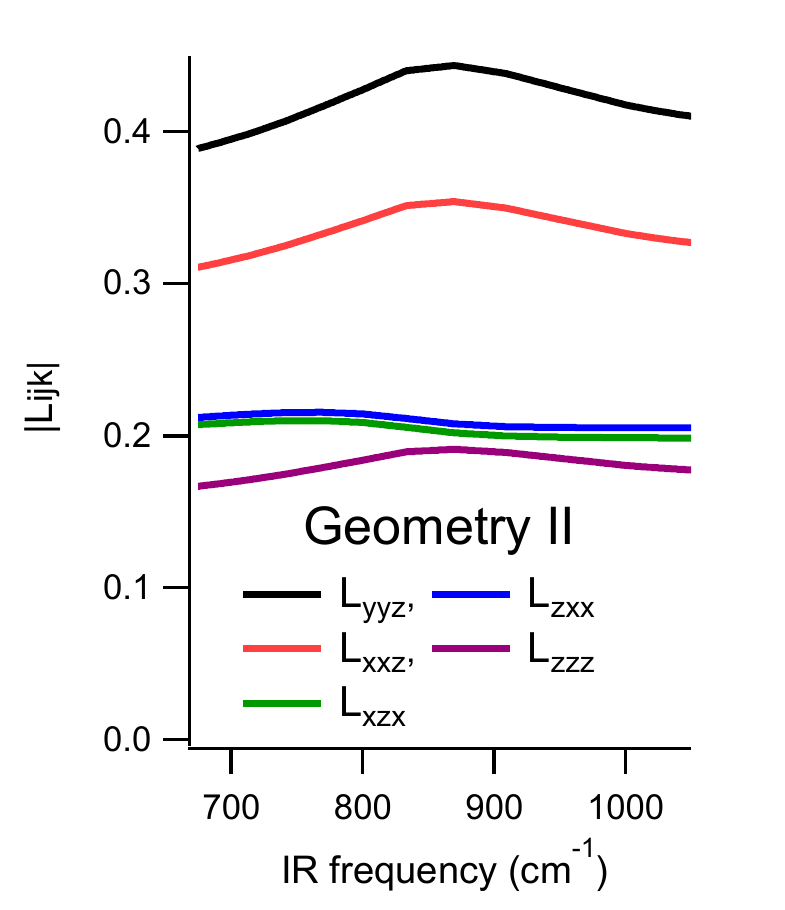} \\
	\multicolumn{2}{c}{\includegraphics[width=0.5\textwidth]{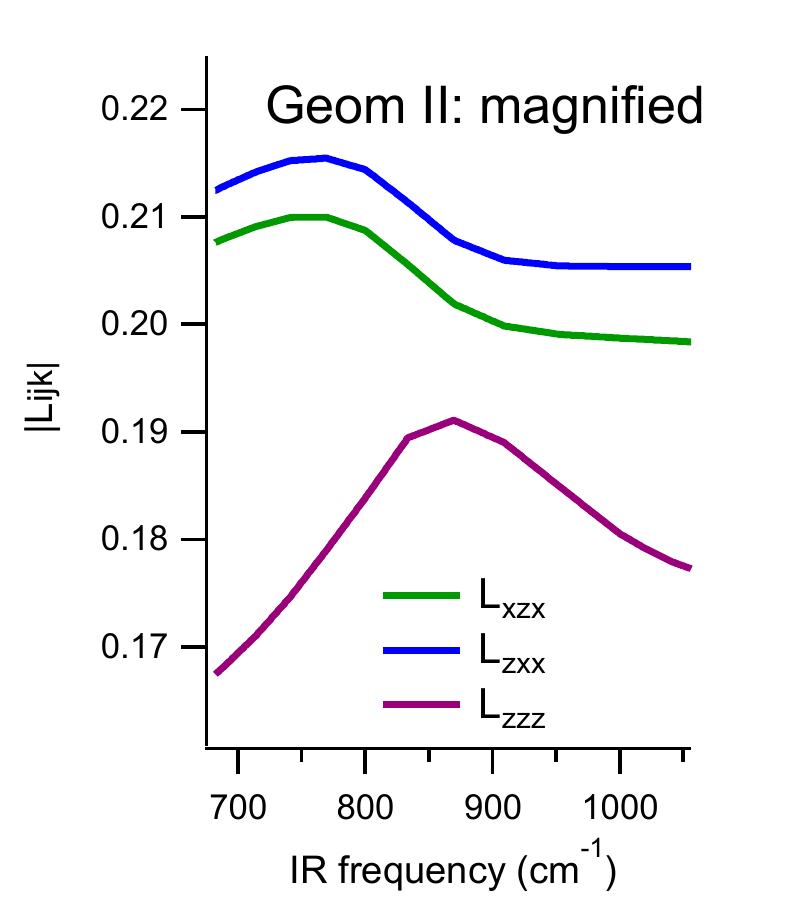}}
	\end{array}$
	\caption{Frequency dependent Fresnel coefficients for the air/H$_{2}$O interface employing the incident angles in geometry I (left panel) and those in geometry II (right and bottom panels). For brevity, the product $L_{ii}(\nu_{\text{\tiny{VSF}}})L_{jj}(\nu_{\text{\tiny{VIS}}})L_{kk}(\nu_{\text{\tiny{IR}}})$ has been abbreviated $L_{ijk}$. }
		\label{f:Lijk}
\end{center}
\end{figure}

The solutions to equations (\ref{e:f_x}), (\ref{e:f_y}) and (\ref{e:f_z}) for the experimental geometries I and II are shown in Figure \ref{f:Lijk}. The large effect of the frequency dependent refractive index of water is apparent in both figures. Clearly one needs to account for this frequency dependence in the linear optical response to extract the true, frequency dependent,  $\chi^{(2)}$. Additionally inspection of the Fresnel coefficients for Geometry 2 in the \textit{ppp} polarization condition highlights that the frequency dependence of the Fresnel coefficients of each of the components of $\chi^{(2)}$ differ dramatically. Clearly accounting for these effects is important in recovering the true sum frequency spectral response. As is clear from the lineshape equations given in the text, spectra measured under the \textit{ppp} polarization condition are a, nontrivial, product of the \textit{xxz}, \textit{xzx}, \textit{zxx} and \textit{zzz} terms.

\subsubsection{Normalization to Account for the IR Spectral Profile}
Our infrared source strongly varies in power as a function of frequency. To account for this variation, as discussed in the manuscript, we measure the $I_{\text{VSF}}$ emitted from a Au mirror at the same location as the sample. Because Au has no resonances in our frequency range of interest $\chi^{(2)} = \chi^{(2)}_{\text{nr}}$. As changes in the refractive index of Au are small over our wavelength ranges of interest, the measured $I_{\text{VSF}}$ is dominated by changes in $I_{\text{IR}}$.

\subsubsection{Normalization to Correct for Changes in Experimental Geometry}
Changing incident beam angles may result in a variety of changes in the experimental set up that make quantitative comparison of sum frequency intensities challenging: spatial and temporal overlap of the two beams may slightly change, the alignment of the detected sum frequency with respect to the spectrometer may change, etc. As mentioned in the text we quantitatively corrected for such unwanted differences in the experimental set up by using z-cut quartz as a reference in the following manner:
\begin{enumerate}
	\item We calculated the frequency dependent Fresnel coefficients, \textit{i.e.} equations \ref{e:f_x}-\ref{e:f_z} and the coherence length in z-cut quartz for our two experimental geometries (as shown in Figures \ref{f:F_zqtz} and \ref{f:wv_zqtz}).
	\item Z-cut quartz has three resonances in our frequency range of interest. Assuming a Lorentzian line shape and employing the fit parameters for each resonance previously published by Shen and coworkers \cite{liu08SI} we calculated the $\chi^{(2)}_{r}$. This result is shown in Figure \ref{f:chir_qtz} and is independent of the incident beam angles. 
	\item Given with this information we calculated the expected change in $I_{\text{sf}}$ with changes in experimental geometry (see Figure \ref{f:Isf_q_calc}). As shown in Figure \ref{f:Isf_q_meas} the calculated trends in resonance amplitude as a function of geometry are fairly well reproduced in experiment (given the finite spectral resolution in our experimental configuration much of the fine spectral detail is lost). 
	\item Following this procedure showed a mismatch between experiment and calculation of $5\%$: it was necessary to divide the $\chi^{(2)}_{ijk}$ extracted from Geometry II by 1.05 to recover the expected value relative to $\chi^{(2)}_{ijk}$ in Geometry I. 
\end{enumerate}

\begin{figure}
\begin{center}
	$\begin{array}{cc}
	\includegraphics[width=0.45\textwidth]{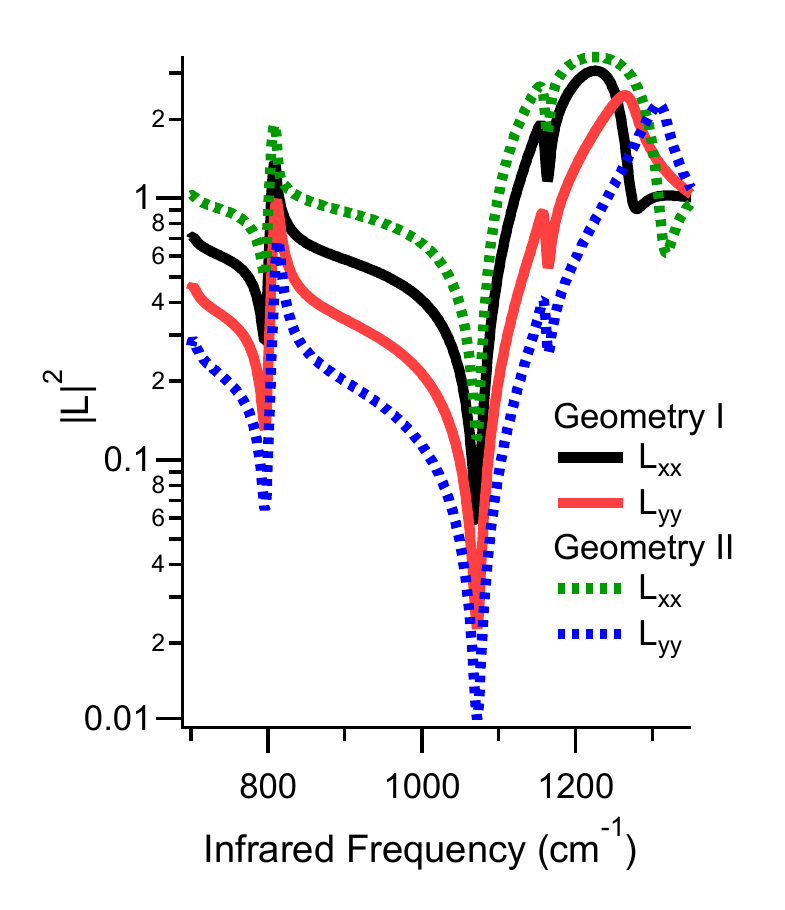} & 
	\includegraphics[width=0.42\textwidth]{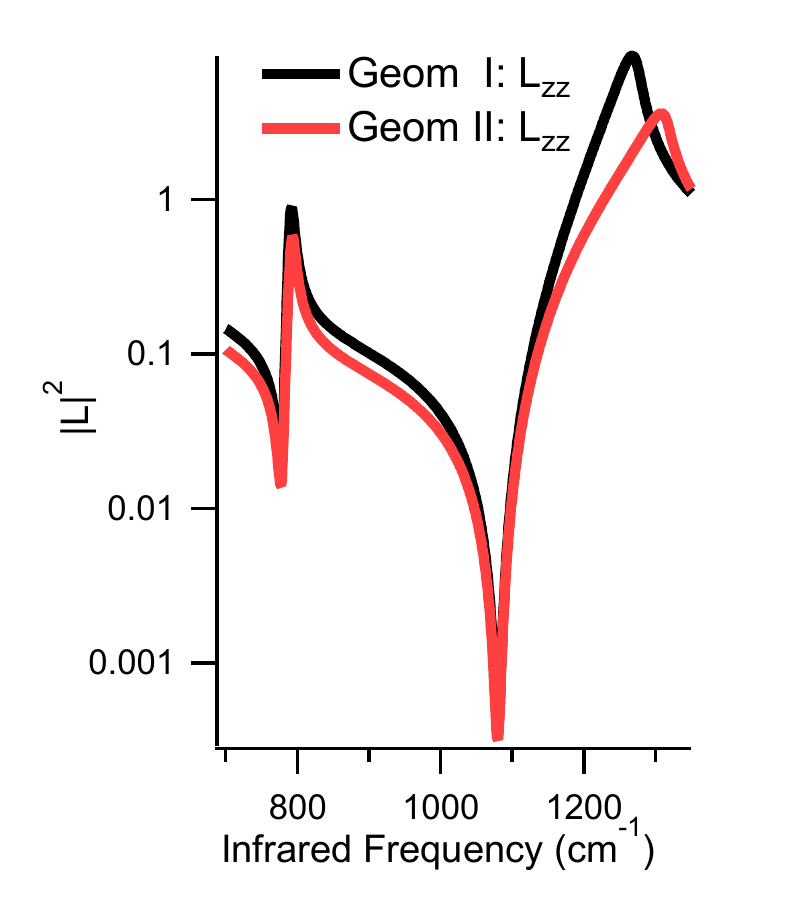}
	\end{array}$
	\caption{Frequency dependent Fresnel coefficients for z-cut quartz under different experimental geometries.}
	\label{f:F_zqtz}
\end{center}
\end{figure}

The error captured in this correction factor is independent of quartz optical properties. Thus we expect it also to apply when changing experimental geometries on the water surface. All reported VSF spectra of the water surface in Geometry II have thus been divided by 1.05. In any case, as is clear from inspection of the data in the paper, all conclusions are insensitive to this correction.

\begin{figure}
\begin{center}
	\includegraphics[width=9cm]{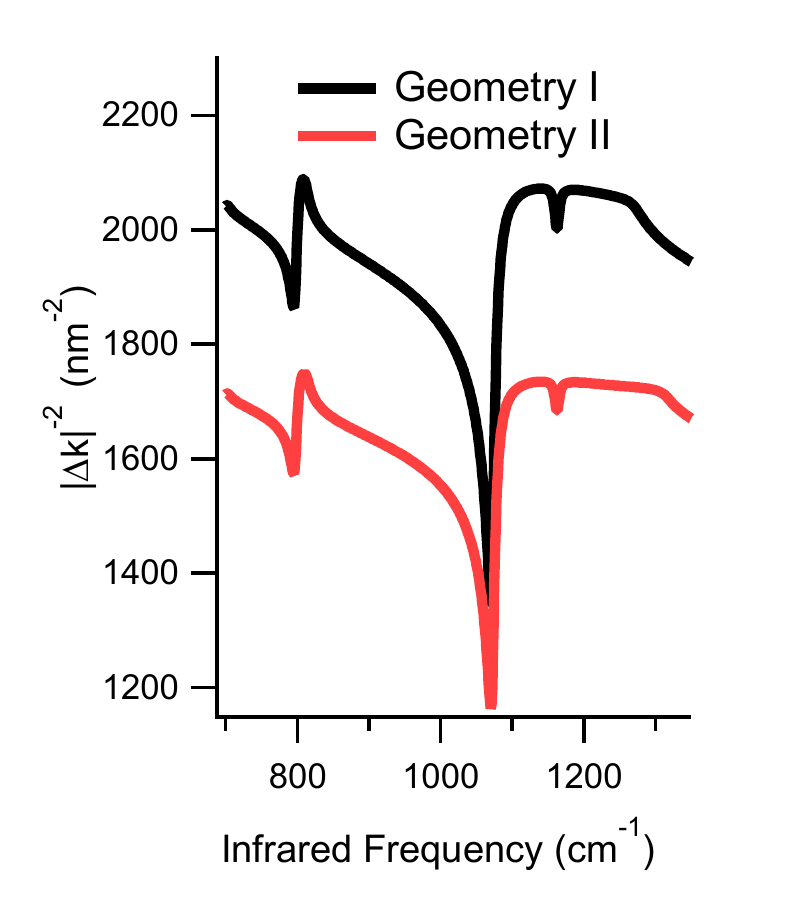}
	\caption{Calculated, frequency dependent, wavevector mismatch of the visible and infrared beams in z-cut quartz with changing experimental geometry.}
	\label{f:wv_zqtz}
\end{center}
\end{figure}

\begin{figure}
	\begin{center}
		\includegraphics[width=9cm]{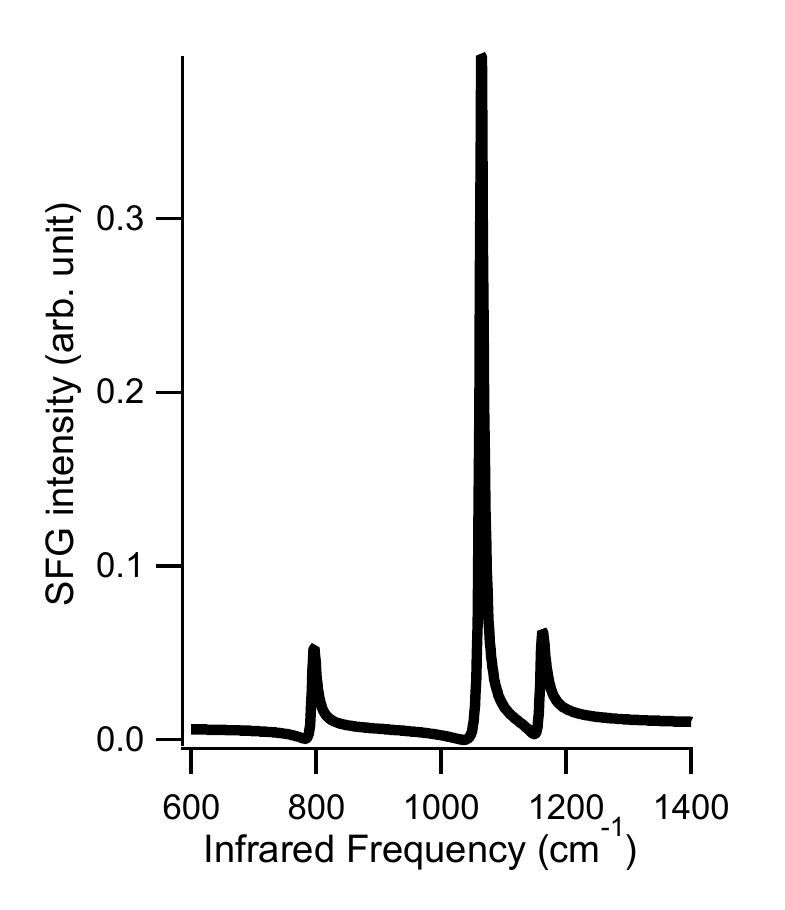}
		\caption{Calculated frequency dependent $\chi^{(2)}_{aaa}$ of quartz using the parameters of Liu and Shen \cite{liu08SI}.}
		\label{f:chir_qtz}
	\end{center}
\end{figure}

\begin{figure}
\begin{center}
	\includegraphics[width = 9cm]{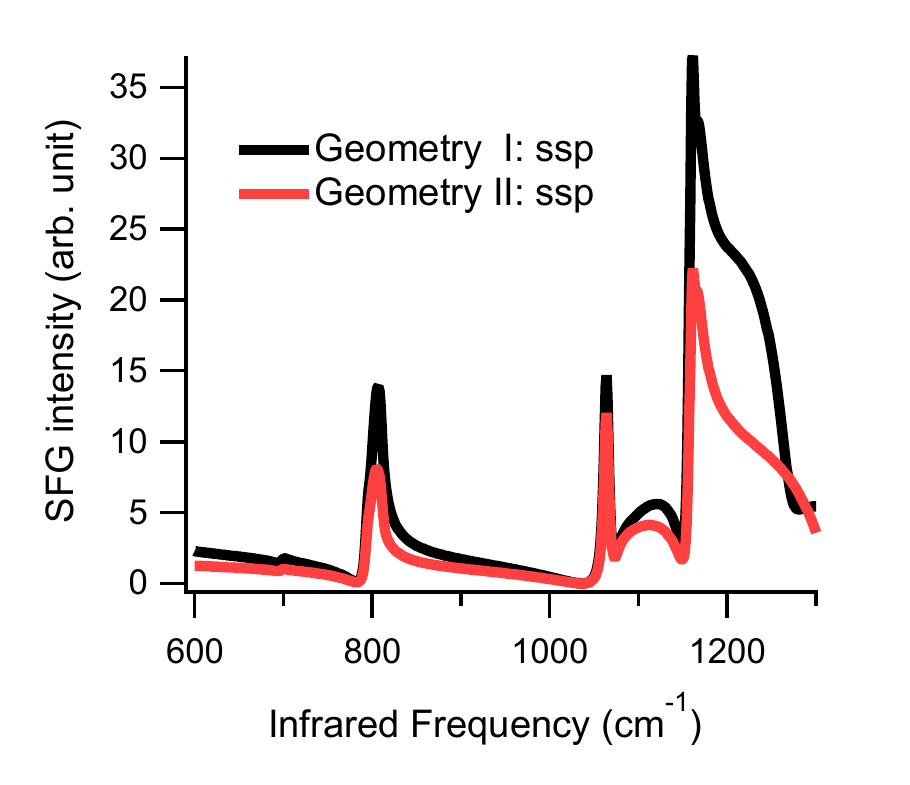}
	\caption{Calculated sum frequency intensity from $\alpha$-quartz as a function of experimental geometry using the results shown in Figures \ref{f:F_zqtz}, \ref{f:wv_zqtz} and \ref{f:chir_qtz}. Note the change in resonance intensity as a function of experimental geometry. This signal assumes that spectral resolution is not limited by the width of the visible pulse or gratings in the spectrometer.}
	\label{f:Isf_q_calc}
\end{center}
\end{figure}

\begin{figure}
\begin{center}
	\includegraphics[width =9cm]{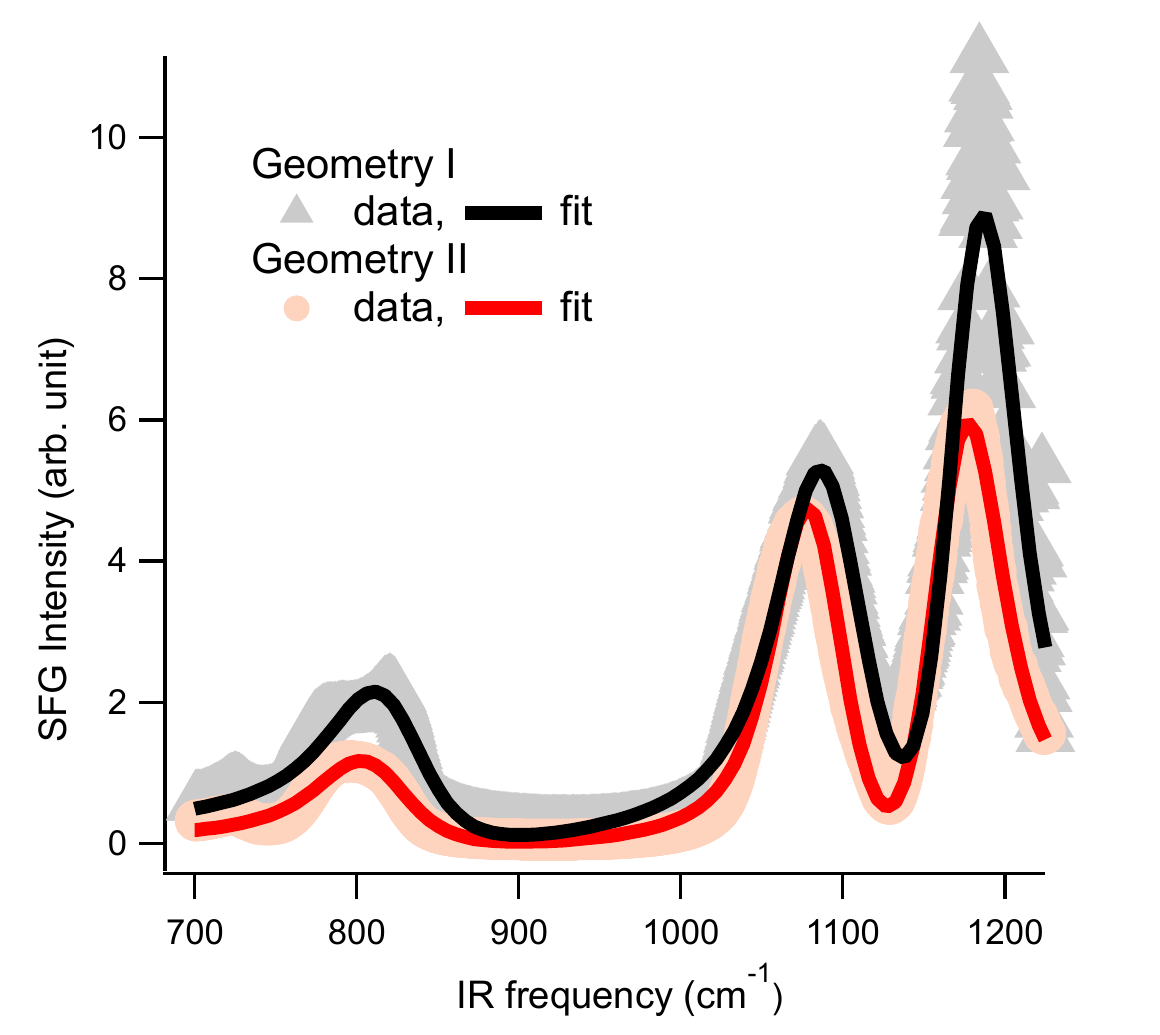}
	\caption{Measured $I_{\text{sf}}$ from $\alpha$-quartz in experimental geometry 1 and 2. All spectra are measured under the \textit{ssp} polarization condition. As is clear from comparison with Figure \ref{f:Isf_q_calc} the change in resonance amplitude with changing beam incident angles is well predicted by the calculation. The measured data is considerably lower resolution owing the spectral width of the visible pulse.}
	\label{f:Isf_q_meas}
\end{center}
\end{figure}

\subsection{Control Experiments}
We determined that our sample was contaminant-free, \textit{i.e.}\ the signal we observe is a property of interfacial \ce{H2O}, by collecting spectra at CH stretch (no CH signal was apparent) and OH stretch (the VSF spectra was quantitatively the same as published reports with our sample preparation procedure) frequencies at the air/\ce{H2O} interface. In addition we also collected spectra at librational frequencies at the air/\ce{D2O} interface that showed no sign of organic contamination. 

In infrared absorption measurements the center frequency of the rotational libration spectral response in D$_{2}$O is red shifted by $\approx 170$ cm$^{-1}$ relative to H$_{2}$O: 500 v.\ 670 cm$^{-1}$ \cite{zel95SI}.  Many prior studies have shown that in the bond stretch frequency range the IR absorption spectra of D$_{2}$O can be quantitatively related to that of H$_{2}$O by accounting for the effect of the change of mass on the force constant. A similar relationship has been shown to exist for the VSF spectra of the OD and OH stretch at the air/D$_{2}$O and air/H$_{2}$O interface respectively \cite{sov08aSI,ton13SI}. It therefore seems reasonable to suggest that, whatever its actual value, the interfacial rotational libration at the air/D$_{2}$O interface should be significantly red-shifted from that at the air/H$_{2}$O. Indeed, as shown in Figure \ref{f:D2O_G2}, this is what we observe and no modes characteristic of, for example, multicarbon organic compounds that might be contaminants, are apparent. 
\begin{figure}
\begin{center}
	\includegraphics[width=9cm]{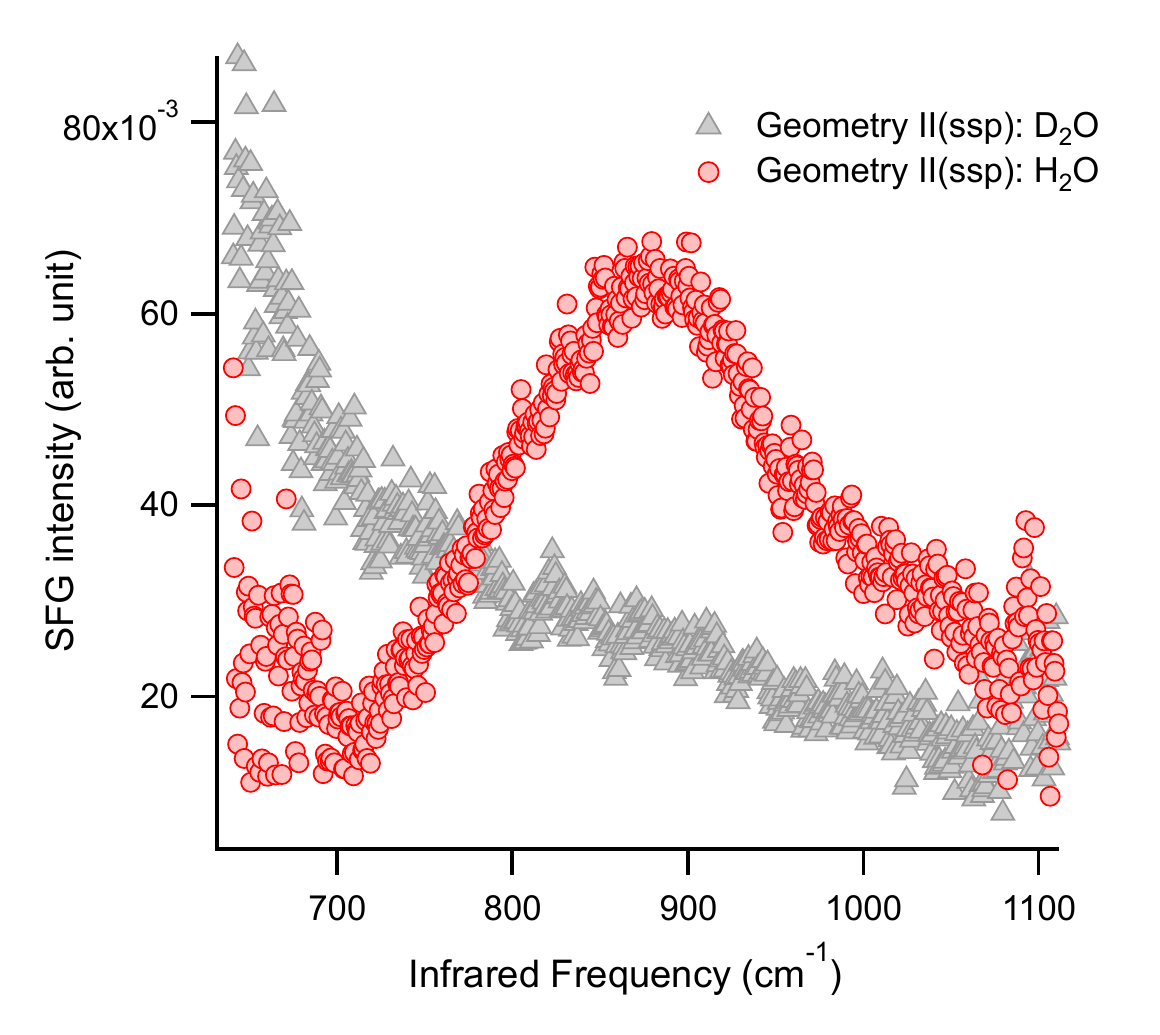}
	\caption{VSF spectral response as a function of infrared frequency for the water librational modes, in Geometry 2 (incident angles of $40^{\circ}$ and $60^{\circ}$ for the visible and infrared) under the \textit{ssp} polarization condition.}
	\label{f:D2O_G2}
\end{center}
\end{figure}

We determined that the signal we observed was a sum frequency signal by its energy (we detected it at the sum of the frequencies of the VIS and IR fiels), its direction (it was detectable only at the expected angle dictated by phase matching), and its power dependence (the measured signal varied linearly with both incident IR and VIS intensities).

\subsection{Details of Quantitative Line Shape Analysis}
\subsubsection{Is a Resonance Required to Fit the Data?}
Because the linear refractive index of water changes significantly over our frequency range of interest it is possible an apparent frequency dependent $I_{\text{\tiny{VSF}}}$ may be the result of a \emph{frequency independent} nonresonant response multiplied by the, \emph{frequency dependent}, Fresnel coefficients. Using the line shape model described in the manuscript, equations (\ref{e:f_x})-(\ref{e:f_z}), treating all nonresonant amplitudes as free parameters and assuming $\chi^{(2)}_{r}=0$  we calculated the frequency dependent $I_{\text{\tiny{VSF}}}$. The result is shown in Figure \ref{f:onlyNR}. Clearly we cannot reproduce the data with $\chi^{(2)}_{r}=0$.
\begin{figure}
\begin{center}
	\includegraphics[width=9cm]{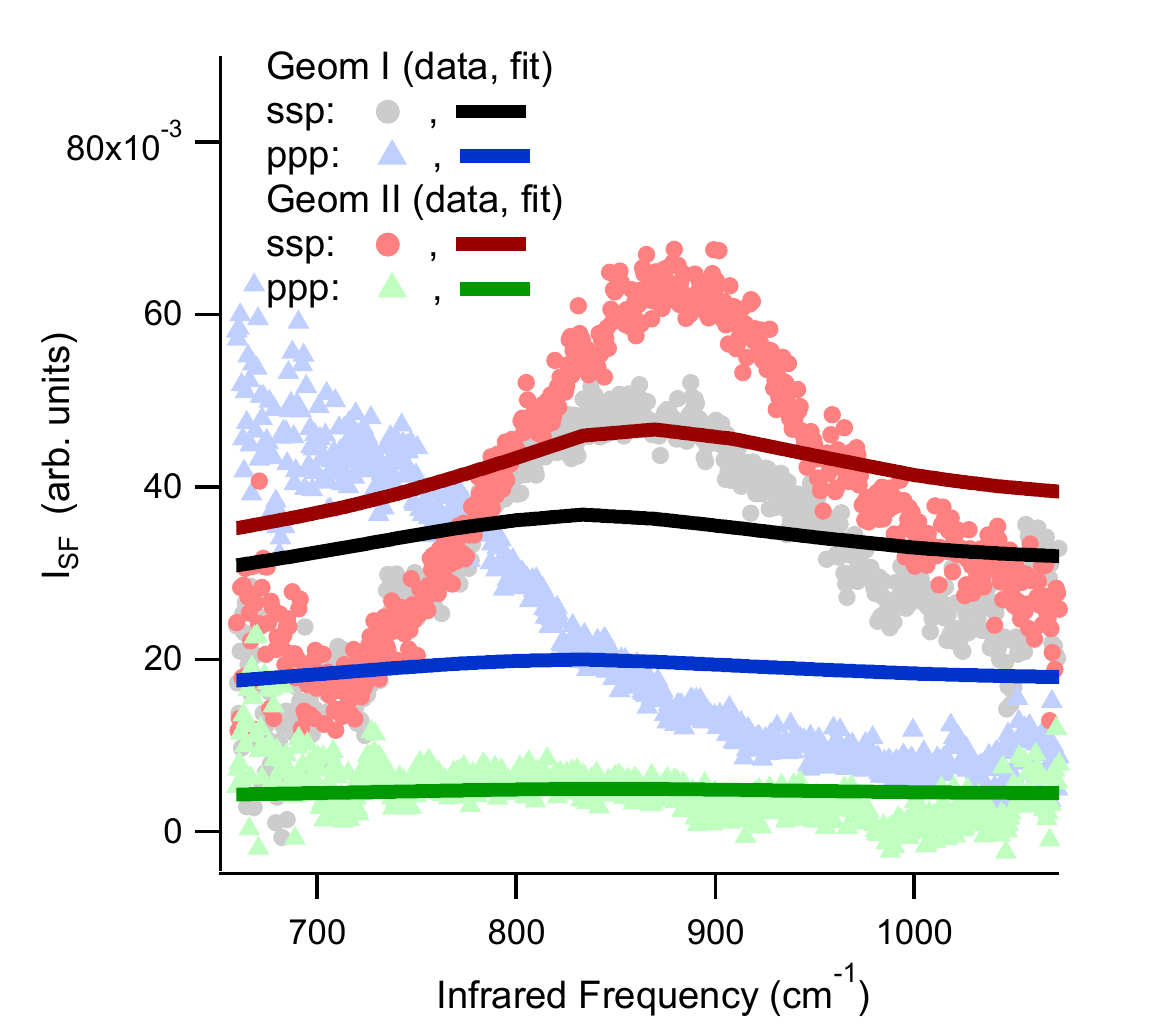}
	\caption{Calculated sum frequency intensity assuming that $\chi^{(2)}_{r}=0$. Clearly the data cannot be reproduced with this assumption.}
	\label{f:onlyNR}
\end{center}
\end{figure}

\subsubsection{Quantifying the Resonant Spectral Response}
Quantifying the observed line shape using the model described in the text is an underdetermined problem: we cannot invert from the data to a unique amplitude (tensor component dependent), line width and center frequency for the observed resonance. However, fitting the data using a number of assumptions about the resonant response illustrates that while the spectral amplitude ascribed to a particular $\chi^{(2)}$ component depends sensitively on assumptions, the center frequency and line width are relatively insensitive. 

In Table \ref{t:data1} we show the results of a global fit to the data from both experimental geometries assuming that $\chi_{nr,xxz}=\chi_{nr,xzx}=\chi_{nr,zxx} =\chi_{nr,zzz}$, $\epsilon_{xxz}=\epsilon_{xzx}=\epsilon_{zxx}=\epsilon_{zzz}$: that the nonresonant amplitude and phase are the same for all components of $\chi^{(2)}_{nr}$ that contribute to the $\chi^{(2)}_{\text{eff,}ppp}$.

\begin{table}[!h]
		\begin{tabular}{rc}
			$\chi_{nr,yyz}$ = & $-0.33\pm0.02$ \\ 
			$\epsilon_{yyz}$ = & $2.07\pm0.06$ \\ 
			$2\Gamma_{\text{vis}}$ = & 37 \\ 
			$\chi_{yyz}=\chi_{xxz}$ = & $75.7\pm5.2$ \\ 
			$\tilde{\nu}$ = & $832\pm3$  \\ 
			$\Gamma$ = & $135\pm3$ \\ 
			$\chi_{nr,xxz}=\chi_{nr,xzx}=\chi_{nr,zxx}$ $=\chi_{nr,zzz}$ = & $0.19\pm0.003$ \\ 
			$\epsilon_{xxz}=\epsilon_{xzx}=\epsilon_{zxx}=\epsilon_{zzz}$ = & $1.99\pm0.01$ \\ 
			$\chi_{xzx}$ = & $-66.4\pm97.5$ \\
			$\chi_{zxx}$ = & $-220\pm138$   \\
			$\chi_{zzz}$ = & $439\pm164$ 
		\end{tabular}
		\caption{Fit parameters for the fits to the data shown in the manuscript. Values of $\epsilon$ are given in radians. Values of $\Gamma$ and $\tilde{\nu}$ are in cm$^{\text{-1}}$, values of $\chi_{nr,ijk}$ are in $10^{-21}\frac{mC\cdot cm}{V^{2}sec}$ and $\chi_{ijk}$ are in $10^{-21}\frac{mC}{V^{2}sec}$.}
		\label{t:data1}
\end{table}

\subsubsection{Can the center frequency of our observed resonance be significantly red shifted?}
As discussed above, unambiguously fitting our observed spectral response is  challenging. Particularly in light of the computational study of Nagata and coworkers \cite{nag13SI}, it is reasonable to ask whether our observations would be consistent with a scenario in which the center frequency of the underlying resonance was at 670 cm$^{-1}$ with a FWHM of 360 cm$^{-1}$ (\textit{i.e.} the line shape of the bulk rotational libration at room temperature determined by Zelsmann and coworkers \cite{zel95SI})  and the intensity maximum observed under the \textit{ssp} polarization condition was shifted to 870 cm$^{-1}$ due to a large nonresonant amplitude and appropriate nonresonant phase. In Figure \ref{f:NR_1}, assuming the indicated resonant center frequency and damping constant, the dependence of the resulting sum frequency emission on nonresonant phase is illustrated. This comparison clearly demonstrates that $\epsilon=0$ is required to move the resulting signal towards our observation. Assuming then, a resonant center frequency of 670 cm$^{\text{-1}}$, a resonant FWHM of 360 cm$^{-1}$ and a nonresonant phase (\textit{i.e.} $\epsilon$) of 0, Figure \ref{f:NR_2} shows the expected signal as a ratio of nonresonant to resonant amplitude (\textit{i.e.} $\chi_{nr,ijk}/\chi_{ijk}$ and compares these results with our data. Clearly it is not possible to shift the resonance sufficiently to reproduce our data and attempting to do so results in spectral distortions, \textit{e.g.} a large apparent baseline, that are not present in the data.
\begin{figure}
\begin{center}
	\includegraphics[width=9cm]{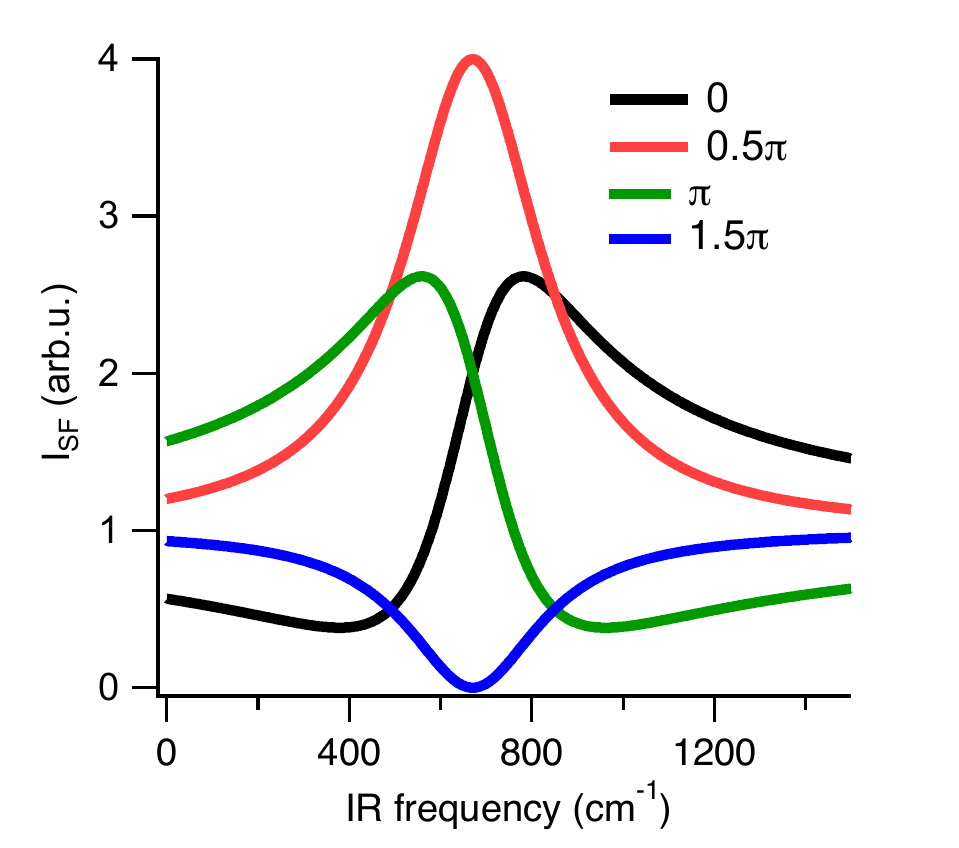}
	\caption{Calculated I$_{sf}$ assuming a single resonance with center frequency 670 cm$^{-1}$ and FWHM of 360 cm$^{\text{-1}}$, $\chi_{nr,ijk}=\chi_{ijk}/\Gamma$ and different values of $\epsilon$ as indicated. Clearly $\epsilon=0$ most efficiently shifts the maximum measured I$_{\text{sf}}$ to higher frequencies.}
	\label{f:NR_1}
\end{center}
\end{figure}

We have analyzed our data assuming $\chi_{nr,ijk} << \chi_{ijk}$ as has been shown to be the case for higher frequency modes of interfacial water. To our knowledge there is no physical mechanism that would invalidate this relationship over our frequency range of interest (750-1050 cm$^{-1}$). Clearly slight deviations from this relationship would make it possible to describe our data with resonance frequencies that are somewhat lower that reported in Table \ref{t:data1}. Nevertheless, as demonstrated in Figures \ref{f:NR_1} and \ref{f:NR_2} such a resonance would still be considerably higher energy than the rotational libration in bulk, thus demonstrating the essential insensitivity of our conclusions to this point. 

\begin{figure}
	\begin{center}
		\includegraphics[width=9cm]{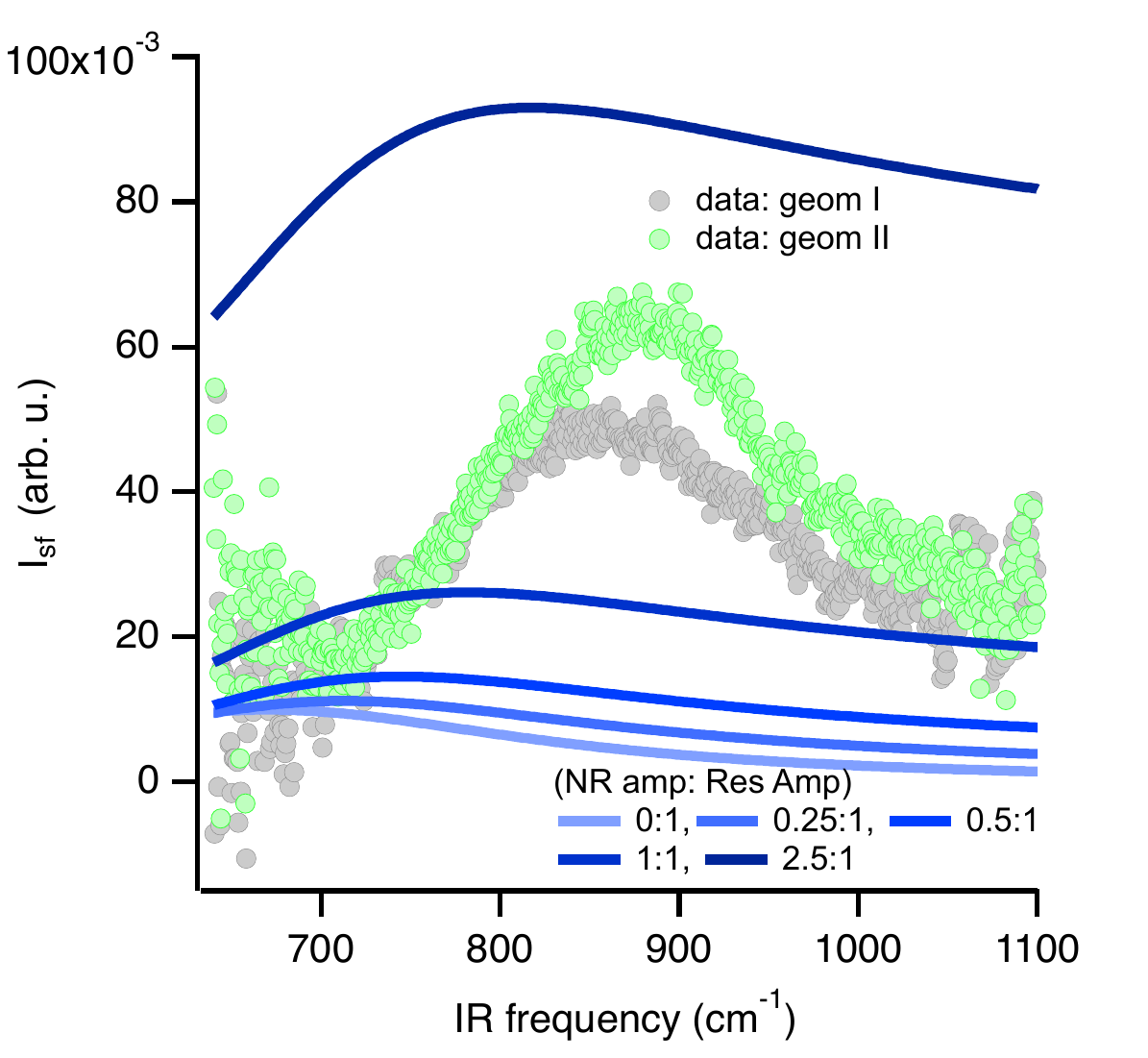}
		\caption{Calculated signal assuming resonant center frequency is 670 cm$^{\text{-1}}$ with a FWHM of 360 cm$^{\text{-1}}$, $\epsilon=0$ and the $\chi_{nr,ijk}/\chi_{ijk}$ ratio indicated. Experimental VSF spectra measured under the \textit{ssp} polarization condition are plotted for comparison as measured in the indicated geometry.}
		\label{f:NR_2}
\end{center}
\end{figure}

\end{document}